\documentclass{llncs}  

\usepackage{graphicx} 
\usepackage[T1]{fontenc}
\usepackage{amsmath}
\usepackage[english]{babel}
\usepackage{array,tabularx}
\usepackage{booktabs} 
\usepackage{makeidx}  % allows for index generation
\usepackage{xparse}
\usepackage{moredefs}
\usepackage{lips}
\usepackage{epstopdf} 
\usepackage{color}
\usepackage{csquotes}
\usepackage{xcolor}
\usepackage{times}
\usepackage[hidelinks]{hyperref}
\usepackage[nolist,nohyperlinks]{acronym}
\usepackage{comment}
\usepackage{paralist}
\usepackage{cleveref}
\usepackage{wrapfig}
\usepackage{multirow}
\usepackage{microtype}
\usepackage{listings}
\usepackage[super]{nth}
\usepackage{natbib}
\usepackage{subfigure}
\usepackage{graphicx}
\usepackage{multirow}

\usepackage{float}
\floatstyle{plaintop}
\restylefloat{table}

\usepackage[style=base,labelfont=bf,labelsep=period,tableposition=top,font=small]{caption}
\makeatletter
\renewcommand\@biblabel[1]{#1.}
\makeatother
\addto\captionsenglish{}

\usepackage{fancyhdr}
\fancypagestyle{WI_footer}{\fancyhf{}\fancyfoot[L]{\color{black}\scriptsize 19th International Conference on Wirtschaftsinformatik\\September 2024, Würzburg, Germany}}

\crefformat{footnote}{#2\footnotemark[#1]#3}
\expandafter\def\expandafter\UrlBreaks\expandafter{\UrlBreaks%  save the current one
  \do\a\do\b\do\c\do\d\do\e\do\f\do\g\do\h\do\i\do\j%
  \do\k\do\l\do\m\do\n\do\o\do\p\do\q\do\r\do\s\do\t%
  \do\u\do\v\do\w\do\x\do\y\do\z\do\A\do\B\do\C\do\D%
  \do\E\do\F\do\G\do\H\do\I\do\J\do\K\do\L\do\M\do\N%
  \do\O\do\P\do\Q\do\R\do\S\do\T\do\U\do\V\do\W\do\X%
  \do\Y\do\Z}

\newcolumntype{L}[1]{>{\raggedright\arraybackslash}p{#1}}   % linksbündig mit Breitenangabe
\newcolumntype{C}[1]{>{\centering\arraybackslash}p{#1}}     % zentriert mit Breitenangabe
\newcolumntype{R}[1]{>{\raggedleft\arraybackslash}p{#1}}    % rechtsbündig mit Breitenangabe

\begin{document}
\frontmatter          % for the preliminaries

\mainmatter              % start of the contributions

\title{Documentation Practices of Artificial Intelligence}

%\subtitle{Research Paper} 

%\author{Blinded\inst{1}}
%\authorrunning{Blinded et al.} % abbreviated author list (for running head)
%\institute{Blinded}

\author{Stefan Arnold\inst{1} \and
Dilara Yesilbas \inst{1} \and
Rene Gröbner \inst{1} \and
\\ Dominik Riedelbauch  \inst{2} \and 
Maik Horn \inst{2} \and
Sven Weinzierl \inst{1}
}

\institute{
    Friedrich-Alexander-Universität Erlangen-Nürnberg, 90403 Nürnberg, Germany \\
    \email{$\{$stefan.st.arnold, dilara.yesilbas, rene.edgar.groebner, sven.weinzierl$\}$@fau.de}
    \and
    Schaeffler Technologies AG \& Co. KG, 91074 Herzogenaurach, Germany \\
    \email{$\{$dominik.riedelbauch, maik.horn$\}$@schaeffler.com}
} 

% -----------------------
% |  Begin of Document  |
% -----------------------
\maketitle
\setcounter{footnote}{0}

% ------------- 
% |  Abstract and Keywords  |
% -------------
\begin{abstract}

Artificial Intelligence (AI) faces persistent challenges in terms of transparency and accountability, which requires rigorous documentation. Through a literature review on documentation practices, we provide an overview of prevailing trends, persistent issues, and the multifaceted interplay of factors influencing the documentation. Our examination of key characteristics such as scope, target audiences, support for multimodality, and level of automation, highlights a dynamic evolution in documentation practices, underscored by a shift towards a more holistic, engaging, and automated documentation. \\


\end{abstract}

%\thispagestyle{WI_footer}

% ------------- 
% |  Content  |
% -------------

\section{Introduction}
\label{sec:introduction}

%The recent evolution of Artificial Intelligence (AI) from research to business requires rigorous documentation standards that encapsulate the operational, technical, legal, and ethical facets \citep{jobin2019global} of its development and deployment.

The rapid transitions of Artificial Intelligence (AI) from academic research to commercial application, particularly in the field of Generative AI (GenAI), requires rigorous documentation standards that encapsulate the operational, technical, legal, and ethical facets \citep{jobin2019global} of its development and deployment.

Serious issues associated with the misuse of AI \citep{tilmes2022disability} prompted research on documentation checklists such as \textit{data statements} \citep{bender2018data}, \textit{model cards} \citep{mitchell2019model}, and \textit{fact sheets} \citep{arnold2019factsheets} to foster accountability. 

%Despite this guidance for AI documentation, the practical application of the checklists often fails to meet the intended effectiveness \citep{bracamonte2023effectiveness, bhat2023aspirations} or there is a lack of willingness for documentation \citep{chang2022understanding}.

Despite numerous checklists to provide guidance for AI documentation, their practical application faces severe reluctance. This can be traced back to a tedious process of gathering information \citep{chang2022understanding} and the perception that checklists fail to meet the intended effectiveness \citep{bracamonte2023effectiveness, bhat2023aspirations}.

To consolidate the understanding and application of AI documentation, we conduct a systematic and comprehensive review \citep{webster2002analyzing, pare2015synthesizing} on documentation practices. By examining the current literature on documentation practices, we aim to delineate the prevailing landscape, identify dominant trends and existing gaps, forecast future directions, and uncover the connections between their \textit{scopes}, \textit{target audiences}, \textit{number of dimensions}, \textit{support for multimodality}, and \textit{degree of automation}. 

\section{Background}
\label{sec:background}

The landscape of AI documentation has expanded significantly to include various forms of transparency and accountability. The current typology of AI documentation checklists predominantly categorizes information at the levels of data \citep{bender2018data, gebru2021datasheets}, models \citep{mitchell2019model}, and systems \citep{arnold2019factsheets}. 

With a shared goal of enhancing transparency, accountability, and ethical practices, each category focuses on different aspects of the AI ecosystem. Documentations at data level focus on dataset provenance, aiming to ensure data quality, fairness, and suitability for training and evaluating AI models. Documentations at model level detail performance metrics and limitations of AI models, and usage guidelines to foster responsible utilization. Documentations at system level aim at recording the broader AI system and its operational context, including its components, interfaces, and compliance with ethical and regulatory standards. These documentations are designed to ensure a responsible AI use across all stages of AI development and deployment. 

However, this three-part typology may not sufficiently address the intricate complexities and diverse stakeholder needs in the AI ecosystem. Critical considerations such as usage implications, support for multimodality, and audience-specific documentation requirements calls for a granulated review of AI documentation. By extending beyond the conventional typology, such an examination offers a more tailored understanding of AI documentation practices for a diverse range of stakeholders.

\section{Related Work}

Several studies on AI documentation illuminate critical areas including adoption \citep{miceli2021documenting, chang2022understanding}, usefulness \citep{bracamonte2023effectiveness}, completeness \citep{bhat2023aspirations}, and trustworthiness \citep{bracamonte2023effectiveness}. These studies consider various scopes, spanning from data \citep{miceli2021documenting}, through models \citep{chang2022understanding, bhat2023aspirations, bracamonte2023effectiveness}, to systems \citep{hind2020experiences}. This breadth highlights the multifaceted nature of AI documentation practices. The methodological approaches in these studies predominantly involve utilizing surveys \citep{bhat2023aspirations, bracamonte2023effectiveness} and interviews \citep{hind2020experiences, miceli2021documenting, chang2022understanding}. This provides a qualitative lens through which the practices and perceptions surrounding AI documentation in various contexts are explored, yielding nuanced insights into its application and impact.

Instead of delving into specific aspects and scopes of AI documentation through qualitative methods, we undertake a quantitative review of the existing literature on AI documentation. Specifically, we aim to map the landscape of AI documentation, providing a broad overview that encapsulates the diversity and depth of documentation practices. This approach allows us to offer a synthesized summary, highlighting key trends, challenges, and opportunities in AI documentation.

\section{Methodology}
\label{sec:methodology}

We aim to explore the landscape and trends within the field of AI documentation. To ensure a clear and thorough communication of our methodological approach, we adhere to the minimum set of items for reporting outlined by \citet{moher2009preferred}.

\paragraph{\textbf{Keywords and Database}}

Since we are  interested in research that falls under the purview of AI documentation, we keyed our search to the term \textit{AI documentation}, broadened by the terms \textit{data cards}, \textit{model cards}, and \textit{system cards}. Using the search terms, we searched the abstract and citation database of \textit{Scopus}, which indexes the digital collections of \textit{IEEE Xplore}, \textit{ACM Digital Library}, and \textit{Elsevier ScienceDirect}. Given its significance in the field of machine learning research, we considered its coverage \citep{mongeon2016journal} adequate for compiling an extensive collection of scholarly articles.

The search terms were required to appear in the title, abstract, or keywords to ensure thoroughness. For quality assurance, only studies from journals and conference proceedings were selected. Publications in any language were included if an English translation existed. The publication year was restricted to begin from 2018, the year \citet{bender2018data} initialized the research on documentation principles.

\paragraph{\textbf{Screening for Retrieval.}}

By applying our search phrases to \textit{Scopus}, we obtained $48$ studies. Through a meticulous screening process, no duplicates were found but $21$ studies were excluded due to irrelevance or insufficient detail on AI documentation practices. 

\paragraph{\textbf{Assessing for Eligibility.}}

To assess the eligibility of the identified studies, we needed a set of criteria for inclusion and exclusion. For inclusion, a study must contribute a checklist for enhancing transparency or accountability of AI. For exclusion, a study must fail to meet basic expectations for rigor and relevance. By matching all studies in full-text against these criteria for inclusion and exclusion, $32$ studies remained. To account for recency, we employed a forward and backward citation analysis to find open-access studies. By applying the same inclusion and exclusion criteria, we identified $7$ additional studies, enlarging our pool of studies eligible for synthesis to a total of $39$.

\paragraph{\textbf{Extraction of Features.}}

Recall that our goal is to map the landscape of AI documentation and uncover connections among their checklists. While the traditional typology is helpful for categorizing AI documentation, it may fall short of capturing the multifaceted needs of the diverse stakeholders engaged in the AI ecosystem. To discover nuanced connections, we meticulously extracted $5$ overarching characteristics crucial for understanding the depth and breadth of AI documentation practices. These characteristics included:

\textit{Scope.} To categorize the diverse landscape of the AI documentation checklists, we applied the conventional typology that subdivides documentation efforts into \textit{data cards}, \textit{model cards}, and \textit{system cards}. To account for guidelines for ethical reporting and communication of AI-generated content, we added \textit{usage cards}, emphasizing responsible dissemination of information. This categorization was instrumental in identifying the specific focus of each paper and understanding how they contribute to the broader conversation about AI transparency and accountability.

\textit{Dimensions.} To capture the breadth of topics covered in the checklists, such as technical specifications and ethical considerations, we quantified the dimensions discussed in each checklist.

\textit{Audience.} To understand the intended audience of each checklist, we specify the audience as decision makers, developers, compliance officers, and end users. This delineation addresses key stakeholders' diverse needs and perspectives. Decision makers address strategic considerations, ensuring AI initiatives align with organizational goals. Developers represent the technical perspective, focusing on implementation and operationalization of AI. Compliance officers encapsulate the regulatory perspective, ensuring AI usage adheres to legal standards and ethical norms, thus safeguarding against compliance risks. End users are considered to capture the applied perspective, emphasizing the practicality and usability of AI systems.

\textit{Multimodality.} This characteristic is encoded as a binary value, indicating whether the documentation covers the integration and interpretation of diverse data sources, such as tabular data, images, or text.

\textit{Level of Automation.} This characteristic describes the level of automation in producing the documentation,  categorizing the process as manual, semi-automated, or fully automated. It provides insights into the efficiency, consistency, and scalability of the documentation process. Manual documentation processes involve human effort in writing, compiling, and updating information. This approach allows for a high degree of precision and personalization but can be time-consuming and prone to inconsistencies due to human error or variability in understanding and interpretation. Semi-automated processes blend human oversight with automated tools to streamline the creation and maintenance of documentation. Automated documentation generates documentation with minimal or no human intervention. Automated systems can quickly produce documentation by analyzing data, model, and system architectures. While this approach offers speed and scalability, it may require sophisticated algorithms to ensure the relevance, accuracy, and comprehensibility of the documentation.

Note that each characteristic was extracted based on explicit mentions or indicators that identify a characteristic. This systematic approach helped us to distill the different facets of AI documentation practices and allowed for a detailed overview of the current landscape and their connections. 

\section{Findings}
\label{sec:findings}

To form an understanding of the nature of AI documentation, we employ descriptive statistics and correlational analysis on the characteristics extracted from the existing body of knowledge. Upon describing and interpreting the distributions and correlations, we synthesize and summarize the most influential documentation practices.

\subsection{Distributions}

\begin{figure}[htp]
    \centering
    \subfigure[Year of Publication]
    {
        \includegraphics[width=0.6\textwidth]{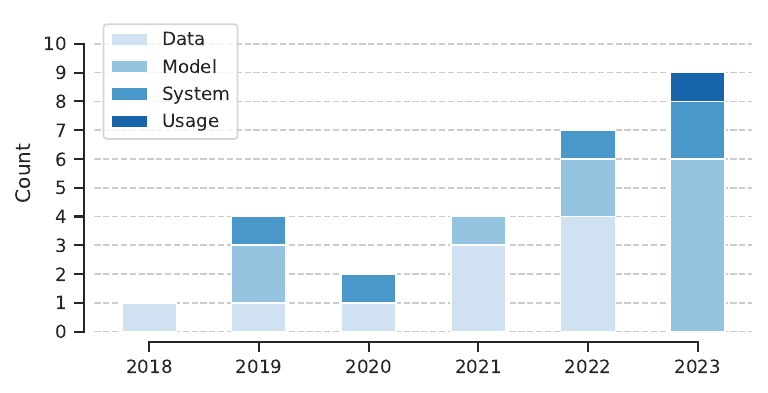}
        \label{fig:year}
    }
    \hfill
    \subfigure[Scope]
    {
        \includegraphics[width=0.3\textwidth]{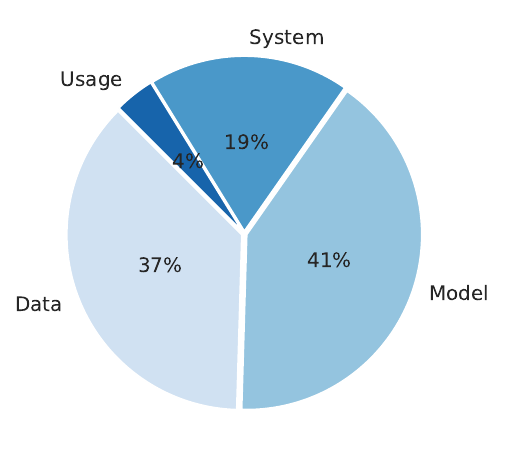}
        \label{fig:scope}
    }
    \caption{Temporal dynamics and distribution of scope.}
    \label{fig:distribution1}
\end{figure}

We commence by investigating the years of publication. Figure \ref{fig:year} presents a temporal distribution of publications on AI cards from 2018 to 2023. The histogram is grouped by the distinct scopes of AI documentation. It unveils a dynamic evolution in documentation practices, with discernible shifts in focus over the observed period, potentially reflecting the community's changing priorities and response to regulatory and ethical imperatives.

% Scope

Figure \ref{fig:scope} provides a breakdown of the scope of AI documentation. The analysis revealed a pronounced emphasis on data and model cards, comprising $41\%$ and $37\%$ of the documentation types respectively, signifying their central role in promoting AI transparency and accountability. The presence of system cards and usage cards, albeit being less prevalent at $19\%$ and $4\%$ respectively, signals burgeoning areas of interest within AI documentation practices. This distribution underscores the AI community's prioritization of data and model transparency, while also pointing to the growing recognition of system-level and usage considerations in AI deployment.

%% Dimensions

\begin{figure}[htp]
    \centering
    \subfigure[Number of Dimensions]
    {
        \includegraphics[width=0.55\textwidth]{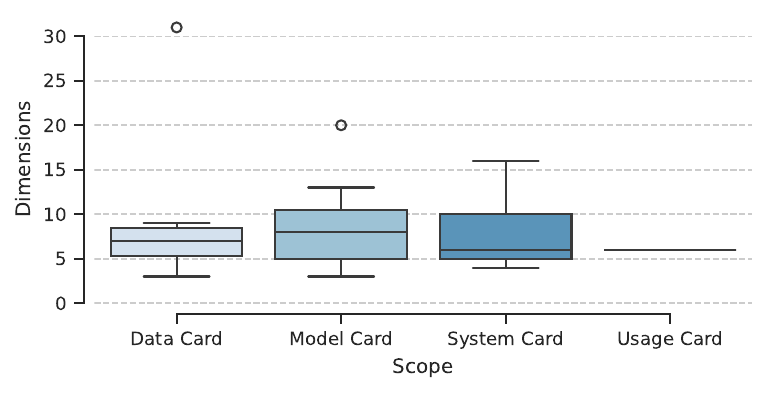}
        \label{fig:dimensions}
    }
    \hfill
    \subfigure[Audience]
    {
        \includegraphics[width=0.35\textwidth]{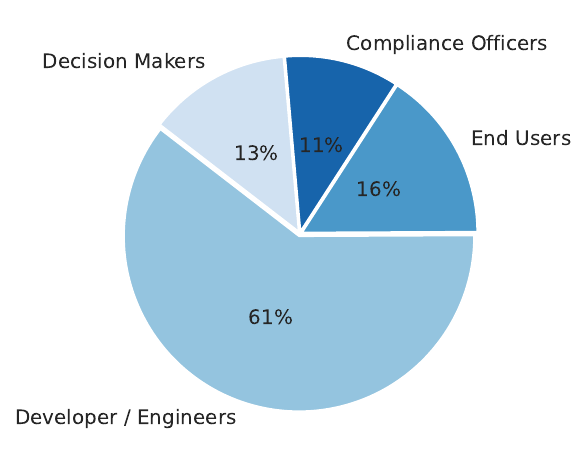}
        \label{fig:audience}
    }
    \caption{Distributions of the dimensionality and target audience.}
    \label{fig:distribution2}
\end{figure}

By quantifying aspects such as operational parameters, technical specifications and ethical consideration, Figure \ref{fig:dimensions} delineates the number of dimensions separated by each scopes of documentation \footnote{Since only one publication deals with documenting AI at usage, there is no trend observable regarding the number of dimensions.}. This distribution underscores the varying emphasis placed on different dimensions within AI documentation, reflecting the nuanced objectives and requirements tailored to each scope (and intended audience).

The average dimension count across all documentation scopes is $7$ with a considerable standard deviation of $3$, reflecting the diverse scope captured within AI documentation practices. Apart from a few outliers, the number of dimensions scales with the scope. This trend suggests that AI documentation is a multifaceted process, with the scope expanding in tandem with the complexity of the AI components being documented. The presence of outliers suggests that while most documentation practices require a relatively small set of dimensions for adequate documentation, there are exceptional cases where specific requirements necessitate a much broader range of documentation criteria. We removed the outliers when interpreting the number of dimensions for each documentation scope. AI cards at data-level have the lowest average dimension count at $5$, but exhibit the highest variability, with some cards requiring documentation across up to $31$ dimensions. This significant spread suggests that while many datasets can be effectively described with a relatively small set of dimensions, the complexity and specificity of certain datasets drive the need for a much more extensive documentation process. At model-level, AI cards show a slightly higher average of $7$ dimensions, indicative of a modest increase in documentation requirements. This could reflect the additional complexity inherent in capturing the nuances of AI models, such as their architecture, training procedure, performance metrics, and operational characteristics. The trend of progressively increasing number of dimensions culminates with AI cards at system-level. These cards have an average dimension count of $8$, reflecting the comprehensive nature of documenting entire AI systems, which encompasses not only the data and models but interfaces, and deployment considerations.

%% Audience

Figure \ref{fig:audience} illustrates the audience distribution. Developers constitute the majority, with $61\%$ of the focus, highlighting the technical orientation of current AI documentation practices. Decision makers and compliance officers are relatively less represented, at $16\%$ and $13\%$ respectively, suggesting a moderate emphasis on governance and regulatory compliance. End users are the least considered audience, at just 11\%, indicating potential areas for increased attention in making AI more understandable to the general public.

Contrasted with the more modest representation of decision makers, end users, and compliance officers, the predominant focus on developers points to potential expanses for broadening the scope of AI documentation to encompass a wider array of stakeholders.

\begin{figure}[htp]
    \centering
    \subfigure[Modality]
    {
        \includegraphics[width=0.6\textwidth]{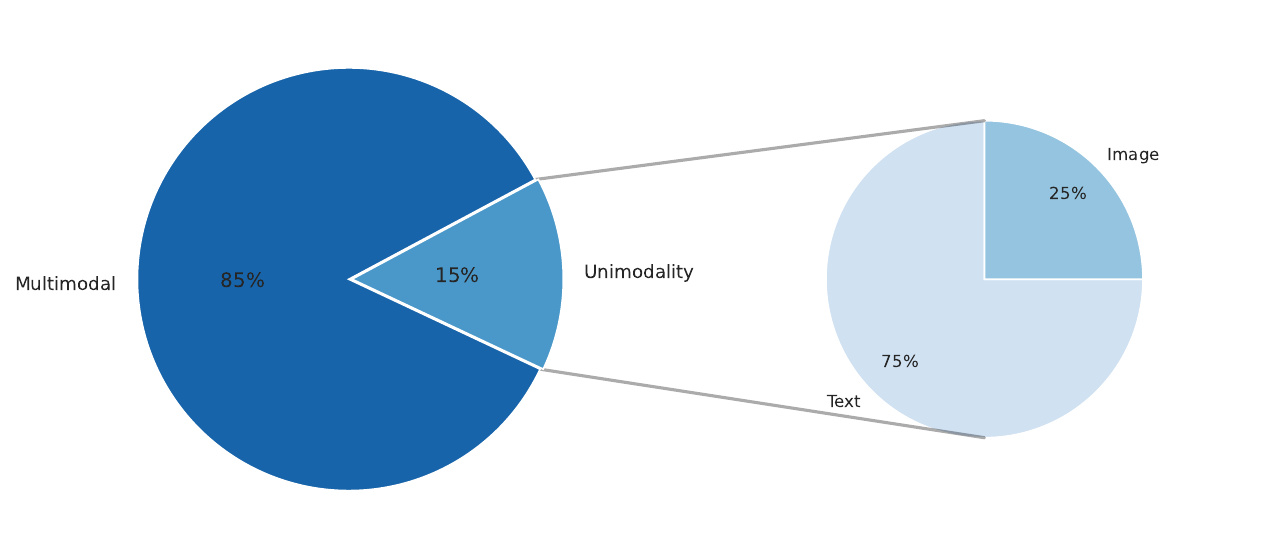}
        \label{fig:modality}
    }
    \hfill
    \subfigure[Level of Automation]
    {
        \includegraphics[width=0.3\textwidth]{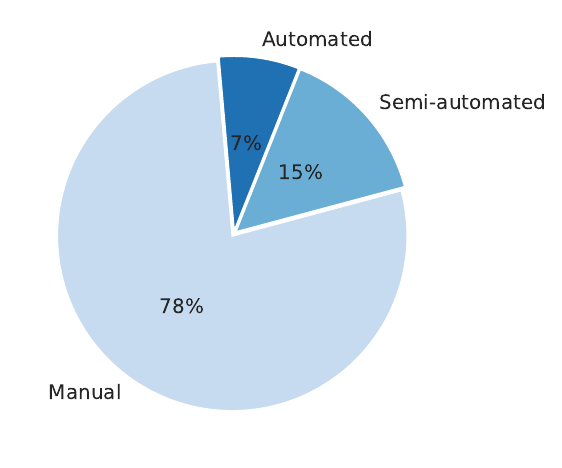}
        \label{fig:automation}
    }
    \caption{Distributions of the multimodality and level of automation.}
    \label{fig:distribution3}
\end{figure}

%% Modality

The distribution of modality support is depicted in Figure \ref{fig:modality}. With few notable exceptions geared towards textual \citep{derczynski2023assessing} and visual sensory \citep{moore2023failurenotes} modalities, a significant share of AI cards accommodates a variety of data formats, making up $85\%$. This multimodal capability underscores the versatility and adaptability of the majority of AI cards. However, a noteworthy fraction of AI cards is unimodal, reflecting considerations unique to their respective modality.

% Automation

Figure \ref{fig:automation} summarizes the levels of automation involved in generating AI documentation checklists. A substantial majority of AI cards, $78\%$, are produced manually, indicating a significant reliance on human input and effort. Another $15\%$ of AI checklists are generated through semi-automated means, which suggests the utilization of tools that can streamline parts of the documentation process while still requiring human oversight. Dimensions that can be automated typically include performance metrics and fairness measures through benchmarks. Only $7\%$ of AI checklists can be produced with minimal or no human intervention. The automated approaches leverage metadata representations for extracting and synthesizing necessary information from AI models and their operational contexts. 

As automation plays a relatively minor role in completing the documentation forms, the current distribution of the level of automation indicates that the AI checklists are still largely dependent on manual efforts. This highlights a lag in the development of tools for automating documentation.

\subsection{Correlations}

\begin{figure}[htp]
    \centering
    \includegraphics[width=0.65\textwidth]{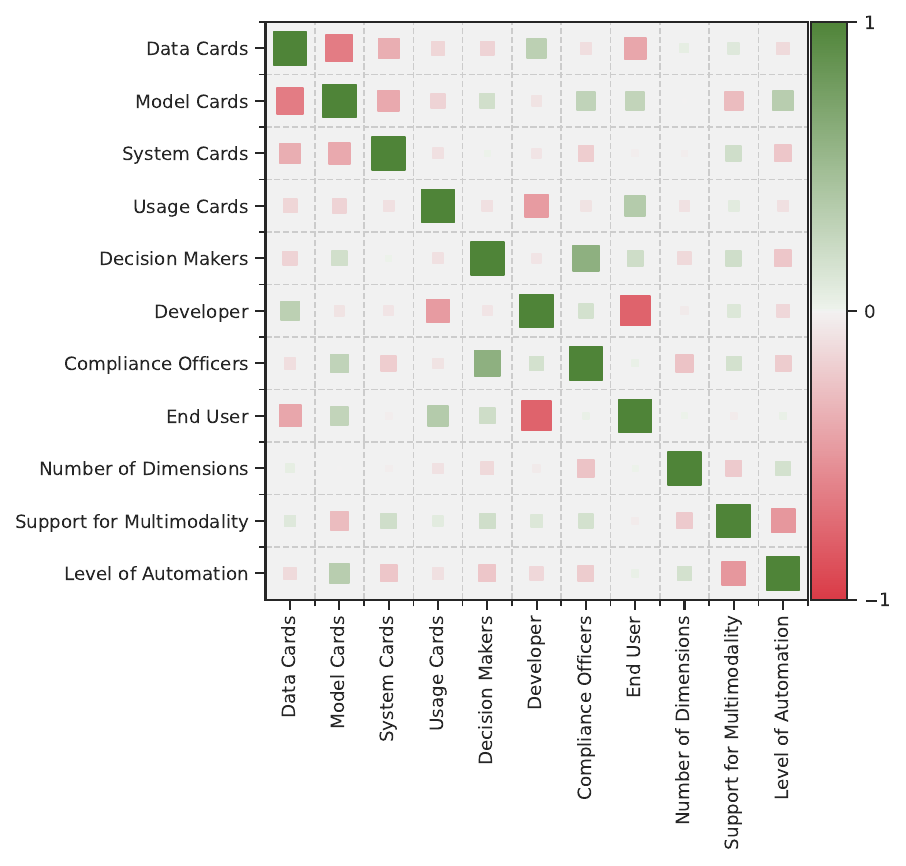}
    \caption{Correlations between characteristics of documentation checklists.}
    \label{fig:correlation}
\end{figure}

To elucidate relationships between documentation checklists, we correlate the extracted characteristics in Figure \ref{fig:correlation}. Several notable correlations warrant detailed discussion.

%By analyzing the intersections between the four different scopes of AI documentation tools with the four main stakeholder groups that have been stated in the majority of underlying papers, you can derive the following assumption: 

The interlink between the documentation scope and their audiences encapsulates the essence of targeted AI documentation. Each card type is tailored to address the informational requisites of its intended stakeholders. The detailed documentation of data cards, encompassing the composition of datasets, aligns mostly with the interests of developers. This suggests that data cards are instrumental for the training and validation of reliable AI systems. Compared to the alignment of data cards with developers, model cards resonate with a diverse audience. This versatility reflects their multifaceted nature, which amalgamate operational insights, technical details, compliance guidelines, and user instructions. Lacking a strong correlation with any audience, the uniformity of system cards underscores their general applicability and suggests that they contain fundamental information pertinent to all stakeholders. Since usage cards are most pertinent to end users, these cards are crucial for ensuring that end users are equipped with knowledge on appropriately and responsibly operating the AI systems. 

Automation is influencing these trends but faces challenges with incorporating multimodality, reflecting the current limitation of automating documentation of AI for diverse data formats. This calls for advanced automation tools handling documentation in various data formats to enrich accessibility of AI documentation.

\subsection{Summarization}

We structure our systematic and comprehensive summary of multimodal checklists according to the scope of documentation at the level of data, model, system, and usage.

\paragraph{\textbf{Data Cards.}} \citet{gebru2021datasheets} advocates a standardized questionnaire for communicating datasets through data sheets, covering details related to composition, collection, and maintenance. The questionnaire encourages careful reflection of the dataset lifecycle. 

\citet{afzal2021data} supplement a semi-automated readiness assessment for datasets. By treating datasets as infrastructure, \citet{hutchinson2021towards} stress the importance of establishing cyclical development practices. \citet{pushkarna2022data} establish documentation that follows the principles of flexibility, modularity, extensibility, and accessibility. This aims to facilitate adoption in practice and at scale. \citet{diaz2022crowdworksheets} accounts for the socio-cultural backgrounds of annotators.

Unlike the checklist format of \citet{gebru2021datasheets} for data documentation, an orthogonal approach presents dataset information in a visually concise and easily digestible format \citep{sun2019mithralabel, holland2020dataset, chmielinski2022dataset}.  

%\citet{ sun2019mithralabel} provides a set of visual widgets delivering information about the dataset among different tasks on the representativeness of minorities, bias, correctness.
%\citet{ holland2020dataset} discuss metadata, provenance, statistics, variables, and dataset characteristics in the form of histograms and heatmaps.

\paragraph{\textbf{Model Cards.}} \citet{mitchell2019model} pioneered model cards for consistent reporting of model characteristics such as intended use, environmental conditions, performance measures, and ethical considerations to promote transparency and accountability. To foster clarity on model functionality and enable better reusability, \citet{gong2023intended} extend model cards by contracts that specify pre-conditions (requirements to be met before using a model) and post-conditions (expected outcomes after using a model).

By incorporating feedback mechanisms, \citet{crisan2022interactive} expand on model cards to make them more interactive and accessible. This facilitates user engagement and emphasizes the role of human-centered design in the documentation process. \citet{adkins2022method} propose actionable instructions to enhance reproducibility. Transitioning to interactive and prescriptive documentation underscore a shift towards a more inclusive approach to documentation, facilitating better accessibility and reproducibility.

Employing a risk-based approach, \citet{bonnier2023towards} advocates for documenting environmental, social, and compliance risks of models throughout their lifecycle in addition to reporting mitigating actions to achieve these considerations. This supports a responsible documentation that aligns with sustainable and ethical standards. 

\citet{seifert2019towards} propose consumer-friendly labels similar to \citet{holland2020dataset} that inform laypersons about the appropriate usage contexts of a model through assessments of their accuracy, fairness, privacy, robustness, and transparency. This enables laypersons to understand when to prefer on human judgment over AI.

To automate the documentation, \citet{singh2023unlocking} compile a dataset to extract aspects such as training configurations, computational resources, benchmark scores, and architectural details. Another direction explores unified representations to reduce human effort in the documentation process. These representations include metadata \citep{li2023metadata} and ontologies \citep{amith2022toward, naja2022using, donald2023towards}. For instance, \citet{li2023metadata} describe models in a way that makes them easily searchable and retrievable from repositories. By leveraging machine-readable metadata, this aims to facilitate the discovery and reuse of models across various domains.

\paragraph{\textbf{System Cards.}} Drawing on supplier's declarations of conformity, \citet{arnold2019factsheets} introduces fact sheets that serve as documents that detail information about purpose, lineage, accuracy, safety, and security of an AI system. This standardized document is designed to be completed by AI service providers for examination by consumers and aims to promote informed decision-making.

Since AI is used in application areas that it was not originally designed, \citet{shadab2020towards} suggest a standardized template that conveys the interoperability between AI systems and the portability of AI components through interface descriptions. This extends the documentation of \citet{arnold2019factsheets} with facets of autonomy.

Unlike other documentation practices, \citet{hupont2023use} emphasize describing the intended purpose and operational use of an AI system rather than technical aspects. By contextualizing use cases, this approach aligns with legal regulations and can help to assess the risk level of an AI system.

\paragraph{\textbf{Usage Cards.}} To facilitate the reporting of AI-generated content, \citet{wahle2023ai} introduce the concept of usage cards. These cards aim to ensure transparency by acknowledging the use of AI models, uphold integrity through the approval of generated content, and establish accountability by clarifying who bears responsibility for the content produced by AI.

\section{Discussion}
\label{sec:discussion}

We highlight three key insights that outline the current landscape and future trajectories for the refinement of documentation practices. Joint efforts by academia and industry are needed to address these insights.

\paragraph{Holistic Documentation.} As AI becomes more complex and integrated, the nature of documentation is evolving to match this complexity. There is a noticeable shift towards more holistic approaches that encompass operational context and system interdependencies.

\paragraph{Engaging Documentation.} There is a trend towards prioritizing interactivity and acknowledging the needs and experiences of stakeholders. This signifies a paradigm shift in how documentation is conceived and utilized. Making documentation more accessible and engaging empowers stakeholders to navigate, query, and understand AI systems more effectively, fostering a culture of informed and responsible AI use.

\paragraph{Automated Documentation.} Despite the advances in AI documentation, the level of automation in producing documentation is notably low. This reliance on manual effort points towards a significant opportunity for improving the documentation processes through automation. Developing sophisticated tools that automate the documentation process is expected to mitigate reluctance and elevate the consistency of documentation.

\section{Conclusion}
\label{sec:conclusion}

Documentation represent a critical mechanism in the realm of an AI ecosystem to foster transparency, accountability, and ethical governance. It acts as a channel through which stakeholders from various facets of the AI lifecycle can gain essential insights about the AI ecosystem. Decision makers are empowered to assess the alignment of the AI ecosystem with operational needs and strategic goals. Developers glean insights into technical intricacies, enabling them to grasp on the reliability of the AI ecosystem. Compliance officers verify adherence to regulatory standards, safeguarding the organization against legal and ethical pitfalls. End users receive guidance on the application and limitations of an AI, which is paramount for informed and responsible usage.

Through a systematic and comprehensive review, we mapped the status and dynamics of documentation practices. We uncovered the multifaceted interplay of factors that influence the documentation. For researchers, this provides a roadmap for future prospects. For practitioners seeking to navigate the landscape of documentation, this study enables them to make informed decisions to enhance transparency and accountability.

%\section{Acknowledgements}

% ----------------
% | Bibliography |
% ----------------

\bibliographystyle{agsm}
\bibliography{literature}

@article{bender2018data,
  title={Data statements for natural language processing: Toward mitigating system bias and enabling better science},
  author={Bender, Emily M and Friedman, Batya},
  journal={Transactions of the Association for Computational Linguistics},
  volume={6},
  pages={587--604},
  year={2018},
  publisher={MIT Press One Rogers Street, Cambridge, MA 02142-1209, USA journals-info~…}
}

@inproceedings{sun2019mithralabel,
  title={Mithralabel: Flexible dataset nutritional labels for responsible data science},
  author={Sun, Chenkai and Asudeh, Abolfazl and Jagadish, HV and Howe, Bill and Stoyanovich, Julia},
  booktitle={Proceedings of the 28th ACM International Conference on Information and Knowledge Management},
  pages={2893--2896},
  year={2019}
}

@article{gebru2021datasheets,
  title={Datasheets for datasets},
  author={Gebru, Timnit and Morgenstern, Jamie and Vecchione, Briana and Vaughan, Jennifer Wortman and Wallach, Hanna and Iii, Hal Daum{\'e} and Crawford, Kate},
  journal={Communications of the ACM},
  volume={64},
  number={12},
  pages={86--92},
  year={2021},
  publisher={ACM New York, NY, USA}
}

@inproceedings{pushkarna2022data,
  title={Data cards: Purposeful and transparent dataset documentation for responsible ai},
  author={Pushkarna, Mahima and Zaldivar, Andrew and Kjartansson, Oddur},
  booktitle={Proceedings of the 2022 ACM Conference on Fairness, Accountability, and Transparency},
  pages={1776--1826},
  year={2022}
}

@inproceedings{hutchinson2021towards,
  title={Towards accountability for machine learning datasets: Practices from software engineering and infrastructure},
  author={Hutchinson, Ben and Smart, Andrew and Hanna, Alex and Denton, Emily and Greer, Christina and Kjartansson, Oddur and Barnes, Parker and Mitchell, Margaret},
  booktitle={Proceedings of the 2021 ACM Conference on Fairness, Accountability, and Transparency},
  pages={560--575},
  year={2021}
}

@inproceedings{afzal2021data,
  title={Data readiness report},
  author={Afzal, Shazia and Rajmohan, C and Kesarwani, Manish and Mehta, Sameep and Patel, Hima},
  booktitle={2021 IEEE International Conference on Smart Data Services (SMDS)},
  pages={42--51},
  year={2021},
  organization={IEEE}
}

@article{holland2020dataset,
  title={The dataset nutrition label},
  author={Holland, Sarah and Hosny, Ahmed and Newman, Sarah and Joseph, Joshua and Chmielinski, Kasia},
  journal={Data Protection and Privacy},
  volume={12},
  number={12},
  pages={1},
  year={2020}
}

@article{chmielinski2022dataset,
  title={The dataset nutrition label (2nd Gen): Leveraging context to mitigate harms in artificial intelligence},
  author={Chmielinski, Kasia S and Newman, Sarah and Taylor, Matt and Joseph, Josh and Thomas, Kemi and Yurkofsky, Jessica and Qiu, Yue Chelsea},
  journal={arXiv preprint arXiv:2201.03954},
  year={2022}
}

@inproceedings{diaz2022crowdworksheets,
  title={Crowdworksheets: Accounting for individual and collective identities underlying crowdsourced dataset annotation},
  author={D{\'\i}az, Mark and Kivlichan, Ian and Rosen, Rachel and Baker, Dylan and Amironesei, Razvan and Prabhakaran, Vinodkumar and Denton, Emily},
  booktitle={Proceedings of the 2022 ACM Conference on Fairness, Accountability, and Transparency},
  pages={2342--2351},
  year={2022}
}

@inproceedings{mitchell2019model,
  title={Model cards for model reporting},
  author={Mitchell, Margaret and Wu, Simone and Zaldivar, Andrew and Barnes, Parker and Vasserman, Lucy and Hutchinson, Ben and Spitzer, Elena and Raji, Inioluwa Deborah and Gebru, Timnit},
  booktitle={Proceedings of the conference on fairness, accountability, and transparency},
  pages={220--229},
  year={2019}
}

@inproceedings{seifert2019towards,
  title={Towards generating consumer labels for machine learning models},
  author={Seifert, Christin and Scherzinger, Stefanie and Wiese, Lena},
  booktitle={2019 IEEE First International Conference on Cognitive Machine Intelligence (CogMI)},
  pages={173--179},
  year={2019},
  organization={IEEE}
}

@inproceedings{crisan2022interactive,
  title={Interactive model cards: A human-centered approach to model documentation},
  author={Crisan, Anamaria and Drouhard, Margaret and Vig, Jesse and Rajani, Nazneen},
  booktitle={Proceedings of the 2022 ACM Conference on Fairness, Accountability, and Transparency},
  pages={427--439},
  year={2022}
}

@inproceedings{adkins2022method,
  title={Method cards for prescriptive machine-learning transparency},
  author={Adkins, David and Alsallakh, Bilal and Cheema, Adeel and Kokhlikyan, Narine and McReynolds, Emily and Mishra, Pushkar and Procope, Chavez and Sawruk, Jeremy and Wang, Erin and Zvyagina, Polina},
  booktitle={Proceedings of the 1st International Conference on AI Engineering: Software Engineering for AI},
  pages={90--100},
  year={2022}
}

@article{gong2023intended,
  title={What is the intended usage context of this model? An exploratory study of pre-trained models on various model repositories},
  author={Gong, Lina and Zhang, Jingxuan and Wei, Mingqiang and Zhang, Haoxiang and Huang, Zhiqiu},
  journal={ACM Transactions on Software Engineering and Methodology},
  volume={32},
  number={3},
  pages={1--57},
  year={2023},
  publisher={ACM New York, NY}
}

@article{li2023metadata,
  title={Metadata Representations for Queryable Repositories of Machine Learning Models},
  author={Li, Ziyu and Kant, Henk and Hai, Rihan and Katsifodimos, Asterios and Brambilla, Marco and Bozzon, Alessandro},
  journal={IEEE Access},
  year={2023},
  publisher={IEEE}
}

@inproceedings{moore2023failurenotes,
  title={fAIlureNotes: Supporting Designers in Understanding the Limits of AI Models for Computer Vision Tasks},
  author={Moore, Steven and Liao, Q Vera and Subramonyam, Hariharan},
  booktitle={Proceedings of the 2023 CHI Conference on Human Factors in Computing Systems},
  pages={1--19},
  year={2023}
}

@inproceedings{bonnier2023towards,
  title={Towards Safe Machine Learning Lifecycles with ESG Model Cards},
  author={Bonnier, Thomas and Bosch, Benjamin},
  booktitle={International Conference on Computer Safety, Reliability, and Security},
  pages={369--381},
  year={2023},
  organization={Springer}
}

@article{singh2023unlocking,
  title={Unlocking Model Insights: A Dataset for Automated Model Card Generation},
  author={Singh, Shruti and Lodwal, Hitesh and Malwat, Husain and Thakur, Rakesh and Singh, Mayank},
  journal={arXiv preprint arXiv:2309.12616},
  year={2023}
}

@article{derczynski2023assessing,
  title={Assessing language model deployment with risk cards},
  author={Derczynski, Leon and Kirk, Hannah Rose and Balachandran, Vidhisha and Kumar, Sachin and Tsvetkov, Yulia and Leiser, MR and Mohammad, Saif},
  journal={arXiv preprint arXiv:2303.18190},
  year={2023}
}

@article{arnold2019factsheets,
  title={FactSheets: Increasing trust in AI services through supplier's declarations of conformity},
  author={Arnold, Matthew and Bellamy, Rachel KE and Hind, Michael and Houde, Stephanie and Mehta, Sameep and Mojsilovi{\'c}, Aleksandra and Nair, Ravi and Ramamurthy, K Natesan and Olteanu, Alexandra and Piorkowski, David and others},
  journal={IBM Journal of Research and Development},
  volume={63},
  number={4/5},
  pages={6--1},
  year={2019},
  publisher={IBM}
}

@inproceedings{shadab2020towards,
  title={Towards an interface description template for reusing ai-enabled systems},
  author={Shadab, Niloofar and Salado, Alejandro},
  booktitle={2020 IEEE International Conference on Systems, Man, and Cybernetics (SMC)},
  pages={2893--2900},
  year={2020},
  organization={IEEE}
}

@article{hupont2023use,
  title={Use case cards: a use case reporting framework inspired by the European AI Act},
  author={Hupont, Isabelle and Fern{\'a}ndez-Llorca, David and Baldassarri, Sandra and G{\'o}mez, Emilia},
  journal={arXiv preprint arXiv:2306.13701},
  year={2023}
}

@article{wahle2023ai,
  title={Ai usage cards: Responsibly reporting ai-generated content},
  author={Wahle, Jan Philip and Ruas, Terry and Mohammad, Saif M and Meuschke, Norman and Gipp, Bela},
  journal={arXiv preprint arXiv:2303.03886},
  year={2023}
}

@article{naja2022using,
  title={Using Knowledge Graphs to Unlock Practical Collection, Integration, and Audit of AI Accountability Information},
  author={Naja, Iman and Markovic, Milan and Edwards, Peter and Pang, Wei and Cottrill, Caitlin and Williams, Rebecca},
  journal={IEEE Access},
  volume={10},
  pages={74383--74411},
  year={2022},
  publisher={IEEE}
}

@article{amith2022toward,
  title={Toward a standard formal semantic representation of the model card report},
  author={Amith, Muhammad Tuan and Cui, Licong and Zhi, Degui and Roberts, Kirk and Jiang, Xiaoqian and Li, Fang and Yu, Evan and Tao, Cui},
  journal={BMC bioinformatics},
  volume={23},
  number={6},
  pages={1--18},
  year={2022},
  publisher={BioMed Central}
}

@inproceedings{donald2023towards,
  title={Towards a Semantic Approach for Linked Dataspace, Model and Data Cards},
  author={Donald, Andy and Galanopoulos, Apostolos and Curry, Edward and Mu{\~n}oz, Emir and Ullah, Ihsan and Waskow, MA and Dabrowski, Maciej and Kalra, Manan},
  booktitle={Companion Proceedings of the ACM Web Conference 2023},
  pages={1468--1473},
  year={2023}
}

@inproceedings{miceli2021documenting,
  title={Documenting computer vision datasets: An invitation to reflexive data practices},
  author={Miceli, Milagros and Yang, Tianling and Naudts, Laurens and Schuessler, Martin and Serbanescu, Diana and Hanna, Alex},
  booktitle={Proceedings of the 2021 ACM Conference on Fairness, Accountability, and Transparency},
  pages={161--172},
  year={2021}
}

@incollection{chang2022understanding,
  title={Understanding Implementation Challenges in Machine Learning Documentation},
  author={Chang, Jiyoo and Custis, Christine},
  booktitle={Equity and Access in Algorithms, Mechanisms, and Optimization},
  pages={1--8},
  year={2022}
}

@inproceedings{bhat2023aspirations,
  title={Aspirations and Practice of ML Model Documentation: Moving the Needle with Nudging and Traceability},
  author={Bhat, Avinash and Coursey, Austin and Hu, Grace and Li, Sixian and Nahar, Nadia and Zhou, Shurui and K{\"a}stner, Christian and Guo, Jin LC},
  booktitle={Proceedings of the 2023 CHI Conference on Human Factors in Computing Systems},
  pages={1--17},
  year={2023}
}

@inproceedings{bracamonte2023effectiveness,
  title={Effectiveness and Information Quality Perception of an AI Model Card: A Study Among Non-Experts},
  author={Bracamonte, Vanessa and Pape, Sebastian and L{\"o}bner, Sascha and Tronnier, Frederic},
  booktitle={2023 20th Annual International Conference on Privacy, Security and Trust (PST)},
  pages={1--7},
  year={2023},
  organization={IEEE}
}

@inproceedings{hind2020experiences,
  title={Experiences with improving the transparency of AI models and services},
  author={Hind, Michael and Houde, Stephanie and Martino, Jacquelyn and Mojsilovic, Aleksandra and Piorkowski, David and Richards, John and Varshney, Kush R},
  booktitle={Extended Abstracts of the 2020 CHI Conference on Human Factors in Computing Systems},
  pages={1--8},
  year={2020}
}

@article{jobin2019global,
  title={The global landscape of AI ethics guidelines},
  author={Jobin, Anna and Ienca, Marcello and Vayena, Effy},
  journal={Nature machine intelligence},
  volume={1},
  number={9},
  pages={389--399},
  year={2019},
  publisher={Nature Publishing Group UK London}
}

@article{moher2009preferred,
  title={Preferred reporting items for systematic reviews and meta-analyses: the PRISMA statement},
  author={Moher, David and Liberati, Alessandro and Tetzlaff, Jennifer and Altman, Douglas G and PRISMA Group*, t},
  journal={Annals of internal medicine},
  volume={151},
  number={4},
  pages={264--269},
  year={2009},
  publisher={American College of Physicians}
}

@article{mongeon2016journal,
  title={The journal coverage of Web of Science and Scopus: a comparative analysis},
  author={Mongeon, Philippe and Paul-Hus, Ad{\`e}le},
  journal={Scientometrics},
  volume={106},
  pages={213--228},
  year={2016},
  publisher={Springer}
}

@article{tilmes2022disability,
  title={Disability, fairness, and algorithmic bias in AI recruitment},
  author={Tilmes, Nicholas},
  journal={Ethics and Information Technology},
  volume={24},
  number={2},
  pages={21},
  year={2022},
  publisher={Springer}
}

@article{webster2002analyzing,
  title={Analyzing the past to prepare for the future: Writing a literature review},
  author={Webster, Jane and Watson, Richard T},
  journal={MIS quarterly},
  pages={xiii--xxiii},
  year={2002},
  publisher={JSTOR}
}

@article{pare2015synthesizing,
  title={Synthesizing information systems knowledge: A typology of literature reviews},
  author={Par{\'e}, Guy and Trudel, Marie-Claude and Jaana, Mirou and Kitsiou, Spyros},
  journal={Information \& management},
  volume={52},
  number={2},
  pages={183--199},
  year={2015},
  publisher={Elsevier}
}

% ---------------------
% |  End of Document  |
% ---------------------

\end{document}